\documentstyle[aps,prb,preprint]{revtex}
\title{Raman Scattering Study of Ba-doped $C_{60}$  with $t_{1g}$ States}
\author{X. H. Chen}
\address{Japan Advanced Institute of Science  and Technology \\
Tatsunokuchi, Ishikawa  923-1292, Japan\\
and Department   of Physics, University of Science and Technology  of China\\
Hefei, Anhui 230026, P. R. China}
\author{S. Taga, and Y. Iwasa}
\address{Japan  Advanced Institute  of  Science  and Technology\\
Tatsunokuchi, Ishikawa 923-1292,  Japan}
\date{December 1, 1998}
\begin{document}
\maketitle

\begin{abstract}
\noindent

Raman spectra are  reported  for Ba doped fullerides, $Ba_xC_{60}$
(x=3, 4, and 6).  The lowest frequency  $H_g$ modes  split into five components  
for $Ba_4C_{60}$ and $Ba_6C_{60}$  even at room temperature, allowing us a 
quantitative  analysis based on the electron-phonon coupling theory.  For the 
superconducting   $Ba_4C_{60}$ ,  the density  of states  at the Fermi energy  
was derived as 7 $eV^{-1}$,  while the total value of electron-phonon coupling
$\lambda$ was found to be 1.0, which is comparable to that of $K_3C_{60}$.
 The tangential $A_g(2)$ mode, which is known as a  sensitive probe for the 
degree  of charge transfer on $C_{60}$ molecule, shows a remarkable shift 
depending on the Ba concentration, being roughly consistent with the full 
charge transfer from Ba to $C_{60}$. An effect of hybridization between Ba and 
$C_{60}$  $\pi$ orbitals  is also discussed.

\end{abstract}
 
\vspace{4mm}

{\bf PACS numbers: 78.30.-j,  72.80.Rj, 74.70.-b}

\section{Introduction}

 Since the discovery of superconductivity  in alkali-metal  doped $C_{60}$,
extensive research on $C_{60}$ and other fullerenes has been carried out
 worldwide, aiming  at understanding the mechanism  for superconductivity and
other related issues in
fullerenes.\cite{varma,schluter,zhang,jishi,mitch,chakravarty,zhou,gunnarsson}  
Most of the theoretical models  assumed that electron-phonon interaction is 
important for superconductivity.\cite{varma,schluter,zhang,jishi}  Based on
the analysis of the linewidths in vibronic spectra
excited either by light (Raman scattering) or by neutrons, the electron-phonon
coupling constant  $\lambda$  for $A_3C_{60}$ (A=alkali metal)  has been 
estimated.  Recently, Winter  and Kuzmany  observed that  the  low frequency
$H_g(1)$ and $H_g(2)$ modes lose all degeneracy and split into five components,
each of which couples differently to the $t_{1u}$  electrons for single crystal
of $K_3C_{60}$  at 80 K.\cite{winter} These  results  revealed that  in the 
superconducting state,  the pairing  is mediated by  phonons with weak or 
intermediate  coupling.
\cite{varma,schluter,mitch,zhou,gunnarsson,winter,guirion,prassides,koller}
The lowest two unoccupied molecular orbitals of $C_{60}$ are both triply
degenerated, having $t_{1u}$ and $t_{1g}$ symmetry. Filling of $t_{1u}$ and
$t_{1g}$ bands with electrons is achieved by intercalation of alkali metals and
alkaline earth metals to $C_{60}$ solids, respectively. Nevertheless,
understanding of the "$t_{1g}$ superconductors" is extremely poor in comparison
with the well known $t_{1u}$ superconductors. Comparison of physical property
in between the $t_{1u}$ and $t_{1g}$ superconductors is of particular interest
from the view point of mechanism of superconductivity. From the $t_{1u}$ symmetry
of the electrons in the conduction band a coupling is only possible to the total
symmetric $A_g$ modes and to the five-fold degenerate $H_g$  modes. While the
   coupling to the $A_g$ mode  is expected to  be weak due to an efficient
   screening effect,  the $H_g$ modes  may  have a significantly strong coupling 
constant  since they allow a Jahn-Teller  mechanism.  A similar  coupling should  
take place in the case of the electrons with $t_{1g}$ symmetry.

Superconductivity of Ba-doped $C_{60}$ was first discovered by Kortan et al,
\cite{kortan}  who claimed that the superconducting phase is bcc $Ba_6C_{60}$.
Recently, Baenitz et al.,\cite{baenitz} on the other hand,  reported that the 
superconducting  phase  is not $Ba_6C_{60}$ but $Ba_4C_{60}$. Very  recently, 
we succeeded to synthesize single phase $Ba_4C_{60}$, and unambiguously 
confirmed that the $Ba_4C_{60}$ is the superconducting phase. In this work, 
we present results of a Raman scattering study  of  single phase $Ba_xC_{60}$
 (x=3, 4 and 6 ) with $t_{1g}$ states. The  results indicate  that the 
electron-phonon interaction is also important for the $t_{1g}$ superconductor, 
particularly in superconducting $Ba_4C_{60}$.  In addition, some amazing  
results were observed, particularly  for the low frequency  $H_g$  modes. 
(1) Raman shift of  the  tangential  $A_g$  mode  for $Ba_6C_{60}$  is much  
larger  than the simple  extrapolation  relationship  between Raman shift and 
charge transfer in alkali  metal  doped  $C_{60}$; while the  radial  $A_g$  
mode  nearly  remains  unchanged with  increasing  charge transfer. (2) The 
Raman scattering behavior  is quite  different among  the three phases  of 
$Ba_3C_{60}$,  $Ba_4C_{60}$ and $Ba_6C_{60}$, especially  for the 
low frequency  $H_g$  modes.  The low frequency  $H_g$  modes  lose all 
degeneracy   and split into five (or four) peaks  at room temperature for the 
$Ba_4C_{60}$ and $Ba_6C_{60}$  samples, each of which couples  differently 
to electrons  with $t_{1g}$  symmetry. The splitting of  low frequency $H_g$ 
modes  into  five components   even at  room temperature is similar to that  
observed  in single crystal of $K_3C_{60}$ at  low temperature  of 80 K.
\cite{winter} This is significant  to understand the splitting  and
to evaluate the electron-phonon coupling constants for all directly coupling
mode, estimating Tc  in Ba-doped $C_{60}$.

\section{Experiment}

Samples of $Ba_xC_{60}$  (x=3, 4 and 6) were synthesized by  reacting
stoichiometric  amount of powers of Ba and $C_{60}$.  A quartz  tube with
mixed powder inside was sealed under high vacuum of about  $2\times 10^{-6}$
torr. The samples of $Ba_3C_{60}$ and $Ba_6C_{60}$ were calcined at 600 $^oC$
for 216 hours with intermediate grindings   of two times. In order to obtain
high quality  $Ba_4C_{60}$  sample, thermal  annealing was carried out at
600 $^oC$ for 1080 hours with  five intermediate  grindings. X-ray
diffraction showed that  all samples  were single phase,
which is also confirmed by the single peak feature of the pentagonal pinch  
$A_g(2)$  mode  in the Raman spectra.

Raman scattering  experiments  were carried out  using the  632.8 nm line of
a He-Ne laser in the Brewster angle backscattering geometry. The scattering
light  was detected with a Dilor xy multichannel spectrometer using a spectral
resolution of  3 $cm^{-1}$.  Decomposition of the spectra  into  individual
lines was made with a peak-fitting routine after a careful subtraction of the
background originating from the laser. In order to obtain  good  Raman
spectra, the samples  were ground  and pressed  into  pellets  with pressure
of about  20 $kg/cm^2$,  which  were sealed  in Pyrex tubes  under a
high vacuum of $10^{-6}$  torr.

\section{Results  and Discussion}

Figure 1 shows room temperature  Raman spectra for the polycrystalline samples
of $Ba_3C_{60}$,  $Ba_4C_{60}$,  and $Ba_6C_{60}$.  For the three samples, 
only one peak  of the pentagonal   pinch $A_g(2)$ mode is observed, providing an
evidence that each sample is in a single phase. These agree fairly well with the 
x-ray diffraction patterns. Interestingly, the three  spectra have different 
strongest lines; they are  $H_g(2)$, $A_g(1)$,  and $A_g(2)$  modes for 
$Ba_3C_{60}$,  $Ba_4C_{60}$,  and $Ba_6C_{60}$, respectively.  
Another thing to be noted is that the half-width  of  all corresponding peaks  
of $Ba_4C_{60}$   is largest among  $Ba_xC_{60}$ (x=3, 4 and 6 ) samples  
except  for the   $A_g(1)$ mode. This result is indicative of an importance 
of electron-phonon coupling in Raman spectrum of  $Ba_4C_{60}$.  Detailed
discussion is given in the following. Also, it is to be pointed out  that the
Raman spectrum  of $Ba_3C_{60}$  sample  is amazingly similar to that of
 $K_6C_{60}$,\cite{zhou1}   suggesting that the electronic states of 
$Ba_3C_{60}$ is similar to that  of $K_6C_{60}$. This is in a fair agreement
with a simple expectation that  $C_{60}$  in both compounds  is hexavalent.

The frequency  of the pentagonal  pinch  mode $ A_g(2)$  decreases  with
 increasing Ba concentration,  similarly to the case of alkali-metal doped 
$C_{60}$.\cite{duclos}   The Raman shift of  the $A_g(2)$ mode  is discussed  
in the following. By contrast, the  frequency  of the radial  $A_g(1)$ mode   
remains  almost  unchanged  with Ba concentration,  being  different from the 
case of $K_xC_{60}$,   where a  slight up-shift  of  the  radial  $A_g(1)$ mode  
was observed.\cite{zhou1}   The low frequency  $H_g$   modes  show dramatic  
changes depending on the Ba concentration.  In particular,  clear  splittings  
are observed for the lowest frequency  $H_g$ modes  of $Ba_4C_{60}$  and 
$Ba_6C_{60}$.  The positions ($\omega$) and halfwidths ($\gamma$) of the 
Raman modes observed are listed in Table I. For comparison,  the  lines for pure
 $C_{60}$  are  included in Table I. In the following, we show detailed analysis 
of $H_g$  modes first, and then, discuss on the $A_g$ modes.

In Fig.2 we show the results of a line-shape  analysis of the Raman spectra
of the $H_g(1)$ modes for $Ba_3C_{60}$, $Ba_4C_{60}$, and $Ba_6C_{60}$ samples.
All modes were fit to a Lorentzian  line shape. For $Ba_3C_{60}$  and
$Ba_6C_{60}$,  a doublet  with Lorentzian  components  is observed, which has 
been observed in $K_6C_{60}$.\cite{zhou1}  However, the $H_g(1)$ mode  
has to be fit with four components for $Ba_4C_{60}$.  This splitting may  be 
attributed to the symmetry lowering due to the orthorhombic structure of this 
material.  A similar behavior  has been  observed in single crystal  $K_3C_{60}$ 
at   80 K,\cite{winter} in which the $H_g(1)$ mode  is split  into five 
components.  Position of the $H_g(1)$ components  for $Ba_4C_{60}$  
sample  is nearly the same as  that  observed in $K_3C_{60}$.

Figure 3 shows the  higher resolution Raman spectra  in the vicinity of 400
$cm^{-1}$  for $Ba_3C_{60}$,  $Ba_4C_{60}$,  and $Ba_6C_{60}$.    
While  the  cubic  $Ba_3C_{60}$  shows a single peak at 432 $cm^{-1}$. 
$H_g(2)$ mode  is  apparently  split  into five components in $Ba_6C_{60}$.  
This splitting of $H_g(2)$ mode in $ Ba_6C_{60}$  is unexpected   since the 
group theoretical  consideration predicts a splitting into two in the space group 
$I_{m\overline{3}}$   ($T_5^h$).  The splitting of the $H_g(2)$ mode might 
suggest a symmetry  lowering which is not detected in the x-ray diffraction.
This type of disagreement between  microscopic  spectroscopy and structural 
analysis was observed in $Rb_3C_{60}$, and still remains  an open question.
\cite{walstedt}   A characteristic  feature of the $H_g(2)$ mode of
$Ba_6C_{60}$ is that  the widths $\gamma$  of  the components  are  almost  
the same except  for the 428 $cm^{-1}$  component. By contrast, the $H_g(2)$ 
mode of $Ba_4C_{60}$ shows  a strong peak at the high frequency edge
associated with a long tailing structure towards lower frequencies.  Linewidth
 and lineshift for the components are clearly related. A theoretical calculation 
shows the electron-phonon coupling constants are very sensitive to the change in 
the normal coordinates, the different components of the mode correspond to the 
different coupling constants.\cite{gunnarsson}   It suggests that the fivefold 
degeneracy  of the mode is lifted and each component couples with a different  
strength to the $t_{1g}$  carriers in $Ba_4C_{60}$.

 Results of a line-shape  analysis of the Raman spectra of the $H_g(3)$ modes  
are shown in Fig.4. A doublet  of $H_g(3)$ is observed for $Ba_3C_{60}$,  
which is ascribed to symmetry-lowering  relative to $C_{60}$  molecules.
\cite{zhou1}  The $H_g(3)$ mode  also displays a splitting  into four both in
$Ba_4C_{60}$ and $Ba_6C_{60}$.  The splitting of the $H_g(3)$  mode in 
$Ba_6C_{60}$  also contradicts with the group theoretical consideration.  It is 
to be pointed out that this anomalous splitting of $Ba_6C_{60}$  $H_g$ 
modes  is observed only in $H_g(2)$ and $H_g(3)$ modes. The other 
$H_g$ modes are singlet  or doublet, being consistent with the group 
theoretical consideration.

In reference  9, Winter and Kuzmany gave several possible  explanations   for
the splitting of the low frequency $H_g$  modes. (1) The splitting  of the 
modes is understood from the merohedral disorder  for the alkali derived
metallic fullerides.\cite{fischer}  This disorder  is of low enough symmetry 
to allow only one dimensional  representations  for all modes. (2) The splitting  
originates  from a Jahn-Teller type interaction.  This interaction can give rise
to  a new vibrational system with rather large number of components, even 
more than five.\cite{auerbach} Also, a contribution  to  the splitting from an
 internal strain between the doped part of the crystal and the undoped part of
 the crystal. In our experiments, the low frequency $H_g$ modes  almost  lose 
all degeneracy  for $Ba_4C_{60}$ and $Ba_6C_{60}$,  and is different from 
that  of $Ba_3C_{60}$  which  is similar to that  of $K_6C_{60}$ at room 
temperature.  In the  case  of $Ba_4C_{60}$, the splitting can be understood 
since the crystal structure is orthorhombic. However, the splitting of 
$Ba_6C_{60}$  is not explained from the crystal structure. Particularly, when 
one considers that $Ba_6C_{60}$  is isostructural  to $K_6C_{60}$, the 
splitting of $H_g(2)$ and $H_g(3)$ modes are considerably anomalous. This  
result might suggest that there exists a symmetry lowing which cannot  be 
detected  by x-ray diffraction. Similar symmetry lowing is observed in the 
NMR spectra of $Rb_3C_{60}$.\cite{walstedt}  The next thing to be pointed
 out is that the splitting is observed even in polycrystalline samples and at
room temperature, in contrast to the case of $K_3C_{60}$. In alkaline-earth-metal 
doped  $C_{60}$,  the local-density  approximation  calculations show  a strong 
hybridization between the alkaline-earth-atom states  and the $C_{60}$  $\pi $ 
states.\cite{saito,erwin} This hybridization, which is absent in the alkali-metal 
doped $C_{60}$, may  play  an essential role for the splitting of low frequency  
$H_g$ modes  at room temperature.

For the components of low frequency $H_g$ modes, a clear relation between
line shift and line broadening is  observed  in $Ba_4C_{60}$,  which is
similar  to that  of single crystal $K_3C_{60}$.  Winter  and Kuzmany  have
pointed out that the electron-phonon  interaction  plays an important  role in
 the broadening and the shift of the lines,  and  they  deduced  electron-phonon 
coupling constants.\cite{winter} The phonon linewidth
broadening  $\gamma_i$  due to the electron-phonon interaction in a metal
can be related to a dimensionless electron phonon coupling constant   
$\lambda_i$ given by \cite{varma,allen}
\begin{equation}
\gamma_i = \frac{1}{g_i}\frac{\pi}{2}N(0)\lambda_i\omega_{bi}^2  
\end{equation}
where N(0) the density of states at the Fermi level per spin and molecule,
and $g_i$ and $\omega_{bi}$  the mode  degeneracy  and the frequency before
any coupling to the electrons, respectively.  The Allen's  formula given above
will be used to derive the coupling  constants for the eight $H_g$ modes.
Frequencies  of pure $C_{60}$  were used as the bare phonon  frequencies.

In the framework of Allen's theory there should be a linear relation of the
form\cite{varma}
\begin{equation}
\gamma =  -\frac{\pi }{2}N(0)\omega _b \Delta \omega   
\end{equation}
between  $\gamma$  the linewidth  and  $\Delta \omega$  the difference between
the bare phonon frequency and the observed frequency.  According to the
experimental values  of the three lowest frequency  $H_g$ modes  in Table I,
the relations between  linewidth  and  frequency shift  is
plotted in Fig.5 for $Ba_4C_{60}$  and $Ba_6C_{60}$ .  The  $\gamma$ and 
$\Delta \omega$  relation for  $Ba_4C_{60}$   is linear  and consistent with that 
expected  from Eq.(2). N(0) can be deduced from the slope. The  density of 
states obtained from the three  $H_g$ modes  are 7 eV$^{-1}$, 4 eV$^{-1}$  and 
3.2 eV$^{-1}$, respectively. The discrepancy may arise from the fact that we
could not use the real bare phonon frequencies for the evaluation.  Geometry
effects may also contribute to the shift and may be different for the modes. For 
$Ba_6C_{60}$,  there exists no relation between the linewidth
and lineshift in Fig.5b. N(0) is much less than 1 eV$^{-1}$ if it were deduced
basing on the relation between linewidth and lineshift in Fig.5b. It suggests
$Ba_6C_{60}$  could not follow electron-phonon coupling  theory.  It further
supports  that $Ba_4C_{60}$ is superconducting phase,  rather than 
$Ba _6C_{60}$.  For the evaluation of the coupling constants  as discussed 
below a value of 7 $eV^{-1}$   is used for N(0). To our knowledge, no N(0) for
 $Ba_4C_{60}$ is available. The calculated N(0) is 4.3 states per eV\cite{saito}  
and an experimental value of 5.6 $eV^{-1}$  was reported for $Ba_6C_{60}$.
\cite{gogia} The  averaged linewidths and the overall coupling constants for
each mode and for all $H_g$ modes for $Ba_4C_{60}$ are listed in Table II,
together with the frequencies for the pure  $C_{60}$.  The averaged linewidths
are directly  evaluated from the  linewiths listed in Table I.
The values for $\lambda_i$   are evaluated  using Eq.(1).

The individual contributions  to the coupling constant  from each $H_g$ mode
are listed in Table II. The  three  lowest  frequency  $H_g$  modes
dominate  the contribution to $\lambda$,  yielding over 70\% of the total
value. Large coupling constants  of the low $H_g$  modes were also observed 
in $K_3C_{60}$.\cite{winter} Within the BCS framework, the superconducting 
transition temperature $T_c$  can been evaluated basing on the experimental 
values  for $\lambda$ by  the McMillan equation
\begin{equation}
T_{c}=\frac{\hbar\omega_{ln}}{1.2k_{B}}exp[\frac{-1.04(1+\lambda)}
{\lambda-\mu^*-0.62\lambda \mu^*}]
\end{equation}
where  $\omega_{ln}$ is  the logarithmic  averaged  phonon frequency, $k_{B}$
is the Boltzmann constant,  and $\mu^*$  is Coulomb  repulsion between 
conduction electrons.   According to the  observed frequencies and the evaluated 
coupling constants, the  $\omega_{ln}$   was determined as  490 $cm^{-1}$.  With
 this value and $\lambda$, the superconducting transition temperature of 7 K can
be evaluated, assuming  the $\mu^*$ value as 0.3, however, which is anomalously
large. The value for $\mu^*$  is much larger than 0.18 in $K_3C_{60}$ in the same 
 way for evaluation of $T_c$. It might suggest a difference between $t_{1u}$ and 
$t_{1g}$  superconductors. To evaluate $T_c$, on the other hand, the logarithmic
  averaged  phonon frequency of 150 $cm^{-1}$ is obtained if the $\mu^*$ is set 
as a reasonable value  0.2. In this case, the phonon frequency is significantly 
smaller than the intramolecular vibration range. Interestingly, the small phonon
energy  associated with superconductivity is also suggested by analysis of another 
$t_{1g}$  superconductor  $A_3Ba_3C_{60}$.\cite{iwasa}

Let us switch to the arguments on the totally symmetric $A_g$ modes. 
Figure 6 shows the Raman shift of the $A_g(2)$  pentagonal pinch
mode  as a function of nominal charge transfer simply derived from the 
chemical formula  for $Ba_xC_{60}$.  In this figure, we plotted the present 
results of $Ba_xC_{60}$,  as well as that of  $K_xC_{60}$  reported
by  Duclos  et al. \cite{duclos}  and the  theoretical   results of Jishi and 
Dresselhaus \cite{jishi1} for comparison.  Since the plots of $Ba_xC_{60}$  
approximately  fall on an extrapolation of $K_xC_{60}$ or theoretical  line, the 
charge transfer value from Ba to $C_{60}$ is almost complete. The molecular 
valences of $Ba_xC_{60}$  (x=3, 4, 6) are regarded as -6, -8, and -12, 
respectively. However, the situation is more complicated than the case of 
alkali doped materials. Several band calculations and experiments
\cite{saito,erwin,niedrig} suggest a strong effect of hybridization of Ba 
and $C_{60}$ orbitals. If this is the case, the net charge  transfer to 
$C_{60}$ is expected to be incomplete. In the present result, however, the 
charge transfer is approximately complete. Moreover, the slope of $Ba_xC_{60}$
 is steeper than that of $K_xC_{60}$ or theory. These results indicate that the 
phonon mode  should be reconsidered in the presence of metal-fullerene 
hybridization.  Especially, there is a difference of 10 $cm^{-1}$  between the 
experimental and theoretical  values for $Ba_6C_{60}$.  The theory of Jishi 
and Dresselhaus  focuses on the mode softening of the  tangential vibrational
 $A_g$ mode  for the alkali-metal  derived fullerides,  the hydridization between
 intercalants  and $C_{60}$  was not  considered.

It can be seen from Table I that the frequency  of the radial $A_g$ mode
for $Ba_{3}C_{60}$  is 506 $cm^{-1}$.   The upshift is as high as 13 
$cm^{-1}$ relative to pure $C_{60}$. But, upon further doping with barium,
 the frequency  nearly  remains  unchanged,  being  different from alkali-metal 
doped $C_{60}$,   which  shows a continuous  hardening of $A_g(1)$ mode as 
a function of alkali metal concentration.\cite{zhou1} The mode-stiffening effect
is due to electrostatic interactions  which produces sufficient stiffening to 
encounter the softening  of the mode  expected on the basis of charge-transfer
 effects.\cite{jishi1,chen} In the  case of Ba derived fullerides, there exists a 
strong hybridization between  the Ba atoms and the  $\pi$-type functions of the 
$C_{60}$ network. This may lead to a decrease in the electrostatic 
 interactions,  so that  the frequency  of the radial $A_{g}$  mode  nearly  
remains  unchanged with  increasing Ba concentration.

\section{Conclusion}

Raman scattering studies of  single phase  $Ba_3C_{60}$, $Ba_4C_{60}$,  and 
$Ba_6C_{60}$ have been carried out. The lowest frequency $H_g$ modes split in
to five components for $Ba_4C_{60}$ and $Ba_6C_{60}$.  A characteristic 
relation  between lineshift and linewidth  is observed  in $Ba_4C_{60}$,  this is 
consistent  with  that  expected by electron-phonon interaction.While 
$Ba_6C_{60}$  does not  exhibit  such behavior. The characteristic  relation  is 
used to evaluate the N(0), the electron-phonon coupling constants are evaluated 
basing  on the Raman  results in the framework of Allen's theory. The  radial 
$A_g$  mode shows a different behavior from alkali derived fullerides, the
frequency  remains unchanged with increasing Ba concentration; the effect  of 
charge transfer on the softening of the tangential $A_g$ mode  is larger  in the 
alkaline-earth metal  doped $C_{60}$ than  in alkali derived  $C_{60}$. These 
discrepancies  may arise from the hybridization between intercalants and $C_{60}$
in alkaline-earth metal doped $C_{60}$.

\section{acknowlegments}

X. H. Chen would like to thank the Inoue Foundation for Science  for financial
support. This work is partly supported by  Grant from the Japan Society  for 
Promotion of Science  (RFTF 96P00104, MPCR-363/96-03262) and from 
the Ministry of Education, Science, Sports, and Culture.

\begin{table}
\caption{ Positions and linewidths  (in parentheses)  for the Raman modes  in
$C_{60}$ and $Ba_xC_{60}$   (x=3, 4 and 6)}
\renewcommand{\arraystretch}{0.93}
\begin{tabular}{c c c c c} 
         &  $C_{60}$   &  $Ba_3C_{60}$   &   $Ba_4C_{60}$   &  $Ba_6C_{60}$   \\
$I_h$  mode   &  $\omega$   ( $\gamma$  )  &  $\omega$   ( $\gamma$  )  &  
$\omega$  ( $\gamma$  )   &   $\omega$   ( $\gamma$  )  \\
        &     ( $cm^{-1}$  )   &  ( $cm^{-1}$  )   &   ( $cm^{-1}$  )   &
        ( $cm^{-1}$  )    \\ \hline
$A_g(1)$    &   493     &   505.9 ( 4.2 )    &  507.2 ( 2.7 )     &  506.5
( 5.0 )     \\
$A_g(2)$    &   1469   &  1430.8 ( 13.0 ) &  1413.4 ( 15.0)  &  1372.5
( 12.1 )  \\
$H_g(1)$    &  270     &   273 ( 5.3 )        &   247.4 ( 18.3 )  &  274.5
( 5.2 )      \\
                   &              &  278.7 ( 4.6 )     &   262.2 ( 1.8 )    &
                   281.8 ( 2.6 )     \\
                   &              &                           &   269.5 ( 10.2 )  
&            \\
                   &              &                           &   279.4 ( 4.9 )    
&              \\
$H_g(2)$   &   431     &   432.3 ( 5.3 )    &   340.8 ( 2.3 )    
&   385.6 ( 4.4 )      \\
                   &              &                           &   381 ( 23.0 )     
&   405.8  ( 2.2 )     \\
                   &              &                           &   396.6 (14.9 )   
&   415.6  ( 2.4 )     \\
                   &              &                           &   413 ( 11.5 )     
&   428 ( 16.8 )       \\
                   &              &                           &   431.9 ( 5.2 )    
&   438.8 ( 2.8 )     \\
$H_g(3)$   &   709     &   648.2 ( 8.5 )    &   621 ( 37.7 )     
&   585.2 ( 4.8 )     \\
                  &               &   681.6  ( 7.8 )   &   651.2 ( 12.3 )  
&  602.1 ( 5.2 )    \\
                  &               &                           &   680.7 (10.5 )     
&   622.3 ( 3.7 )     \\
                  &               &                           &   694 ( 9.5 )      
&    651.8 ( 12.0 )   \\
$H_g(4)$  &   773      &  760.7 ( 8.4 )     &   751.9 (11.0 )   
&   732.5 ( 8.5 )      \\
$H_g(5)$  &   1099    &  1091.7 ( 18.5)   &   1090.6 ( 11.8 ) 
&  1082 ( 6.0 )       \\
                  &               &  1117.3 ( 12.8 )  &   1117( 12.0 )   
&                             \\
$H_g(6)$  &   1248    &   1227.6 ( 16.1)  &    1215.4 ( 9.4 )  
&  1224( 26 )         \\
                 &                &                            &    1234.1 ( 13.8 ) 
&                            \\
$H_g(7)$  &  1426     &   1322.1 (42.6)   &     1322 ( 48.6 )    
&                           \\
                  &               &   1388.1 ( 26.7 ) &     1381 ( 32.1 )  
&                            \\
$H_g(8)$  &  1573     &    1474.4 ( 26.1 ) &     1461.7 ( 51.0 ) 
&   1437 ( 25.0 )  \\
\end{tabular}

\newpage

\caption{ Positions,  averaged linewidths and electron-phonon coupling constants
normalized to the   density  of states at the Fermi energy for eight
fivefold degenerate  Hg  modes  for $Ba_4C_{60}$   sample.}
\begin{tabular}{c c c c} 
Modes  &  $\omega$  ( $cm^{-1}$ )   &   $\overline{\gamma}$ ( $cm^{-1}$ )    &
$\lambda /N(E_F)$  \\
$H_g(1)$  &  270    &   8.8    &   0.062   \\
$H_g(2)$  &  431    &   11.4   &  0.032   \\
$H_g(3)$  &  709    &   17.4   &  0.018   \\
$H_g(4)$  &  773    &   11.0   &  0.009   \\
$H_g(5)$  &  1099  &   11.9   &  0.005   \\
$H_g(6)$  & 1248   &   11.6   &  0.004   \\
$H_g(7)$  &  1426  &  40.0   &  0.010   \\
$H_g(8)$  &  1573  &  51.0   &  0.011   \\
$\Sigma $  &            &            &  0.151   \\   
\end{tabular}
\end{table}

\noindent
{\bf FIGURE CAPTIONS} \\

\noindent
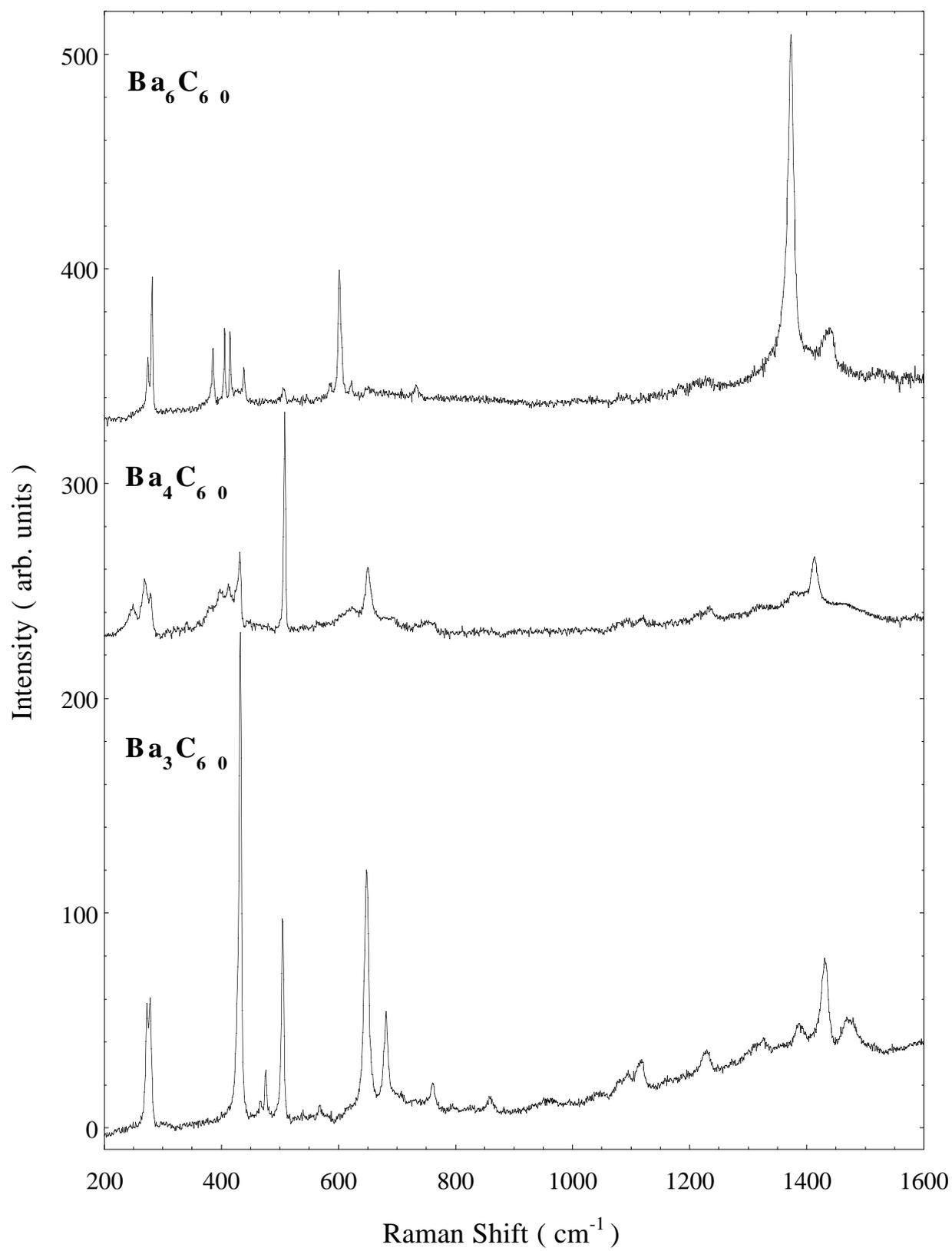
Figure 1:

Room temperature  Raman spectra of $Ba_3C_{60}$,  $Ba_4C_{60}$,  and
$Ba_6C_{60}$.\\

\noindent
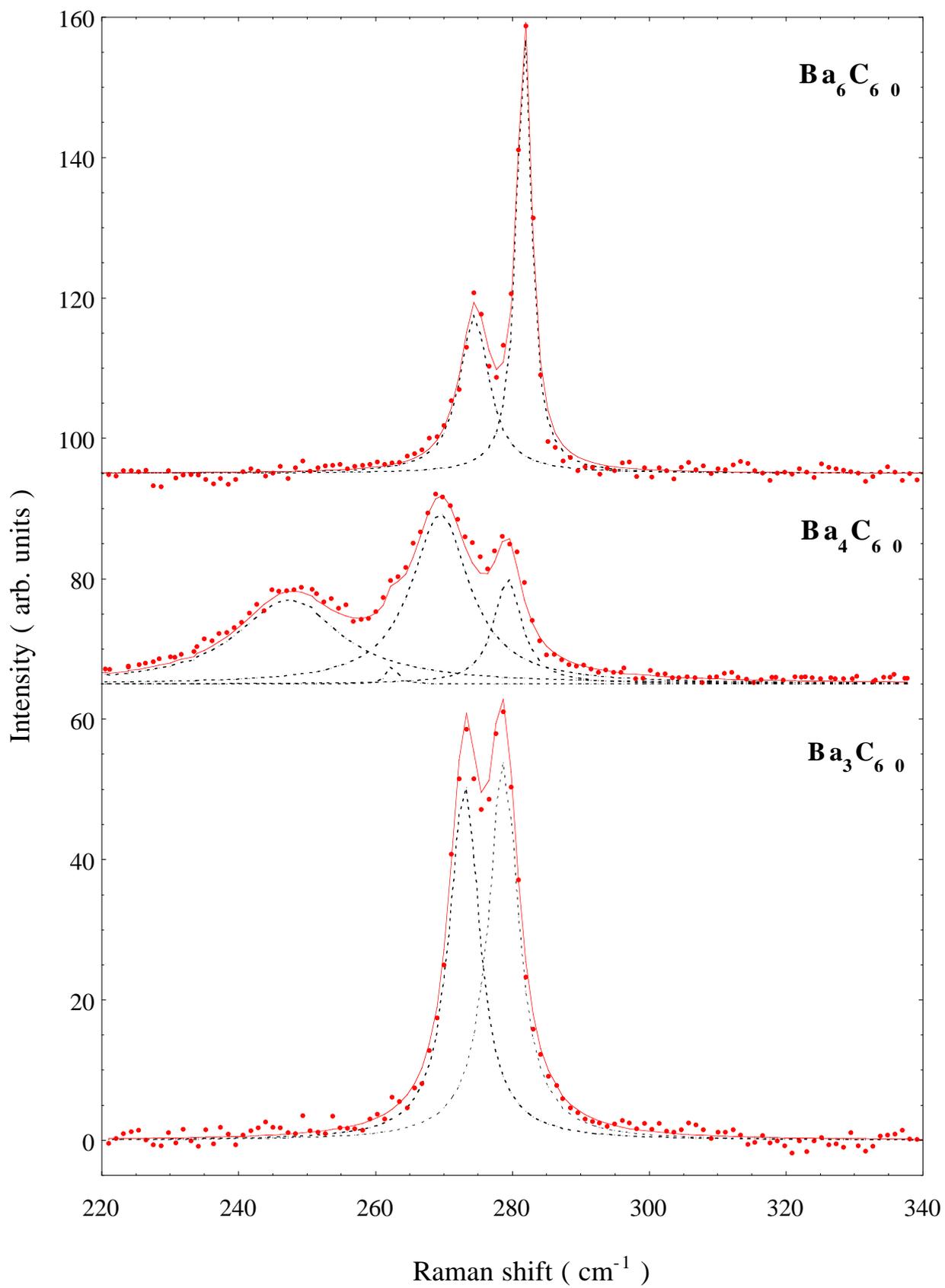
Figure 2:

Raman spectra of the $H_g(1)$ mode  for $Ba_3C_{60}$,  $Ba_4C_{60}$,  
and  $Ba_6C_{60}$.  The  dash lines are computer  fits for the individual
components,  which add up to the full line on the top of the experimental
results.\\

\noindent
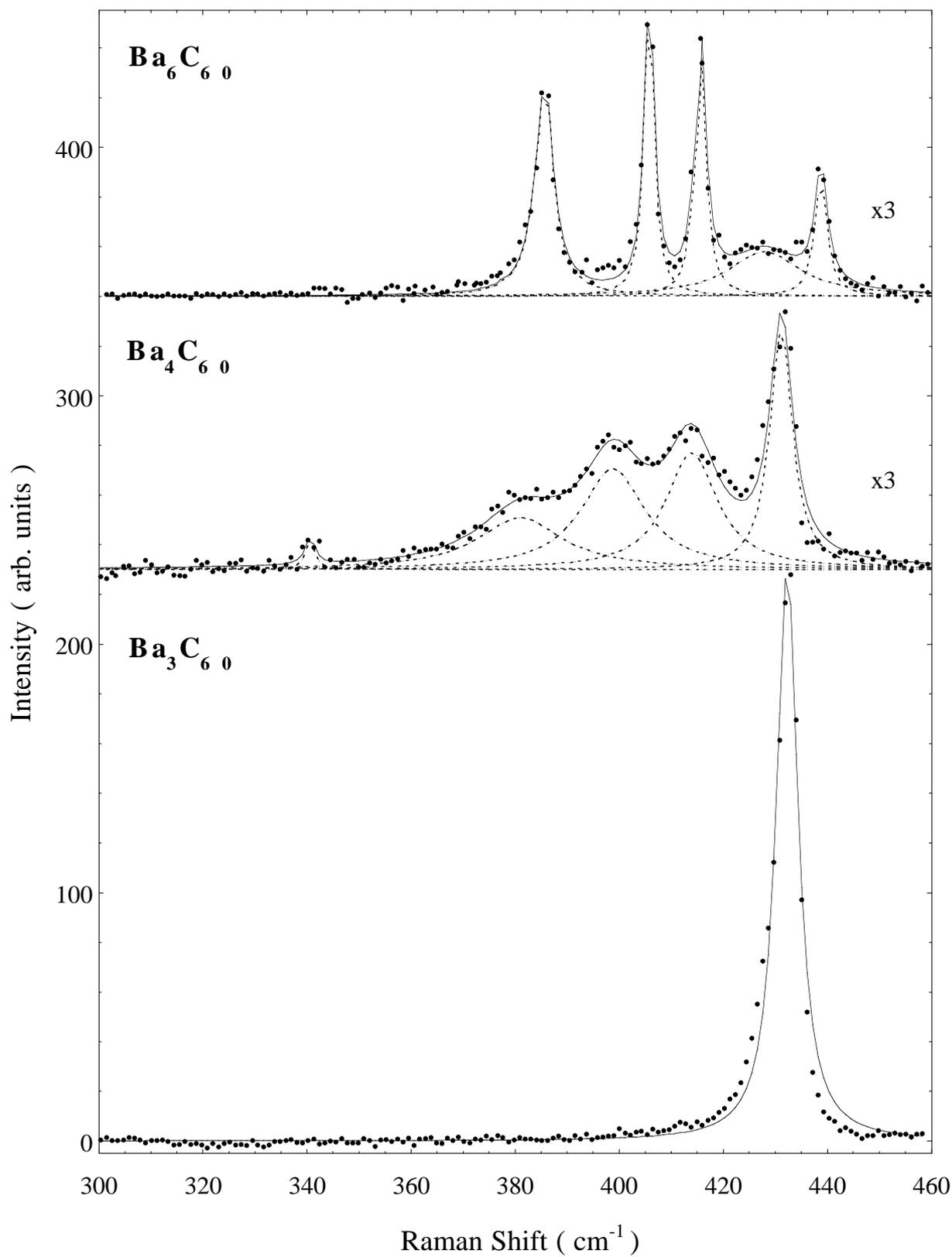
Figure 3:

Raman spectra of the $H_g(2)$ mode  for $Ba_3C_{60}$,  $Ba_4C_{60}$,  
and  $Ba_6C_{60}$.   The dash lines and full line have the same  meaning  as in
Fig.2\\

\noindent
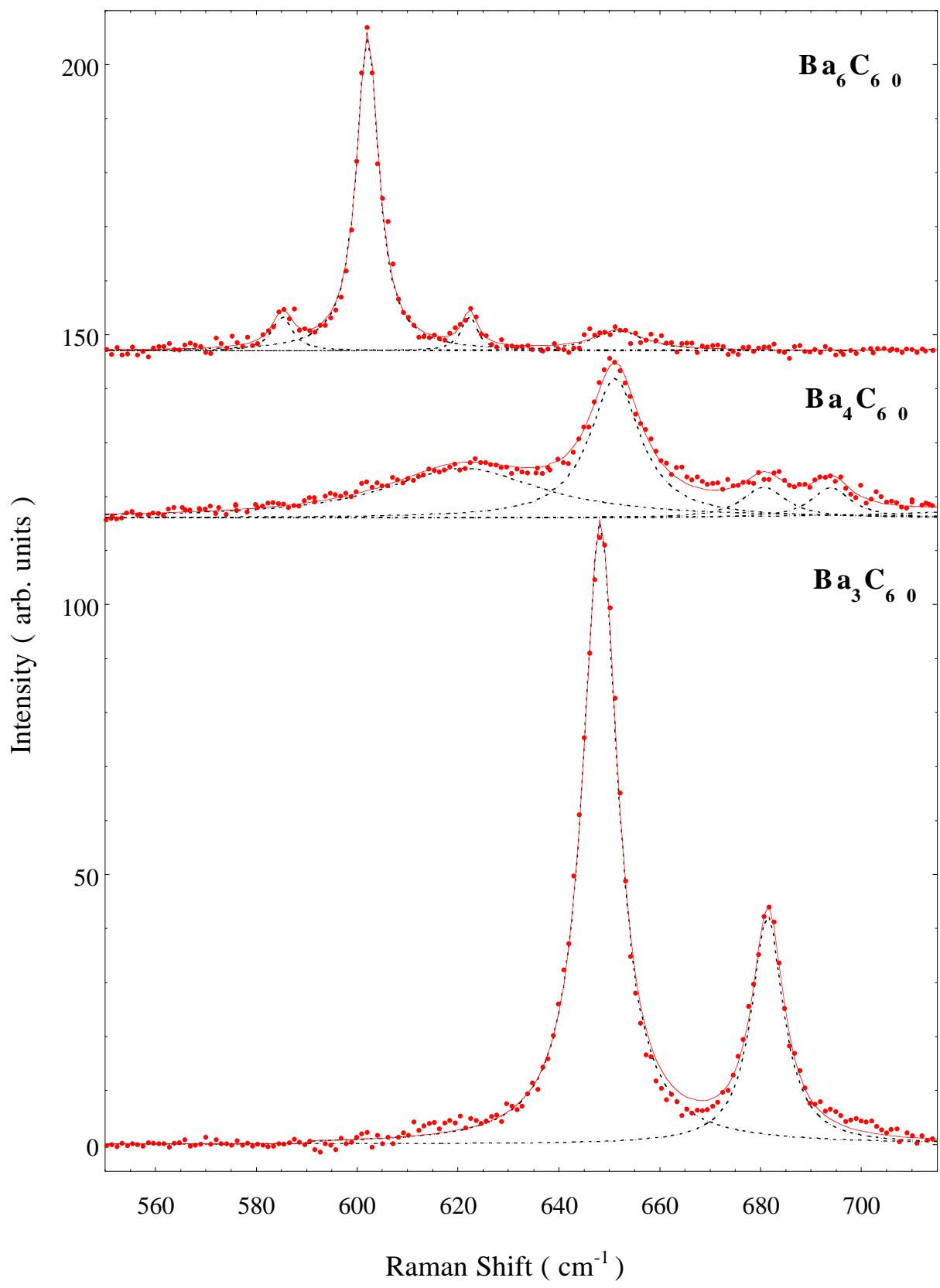
Figure 4:

Raman spectra of the $H_g(3)$ mode  for $Ba_3C_{60}$,  $Ba_4C_{60}$,  and
$Ba_6C_{60}$. The dash lines and full line have the same  meaning  as in Fig.2\\

\noindent
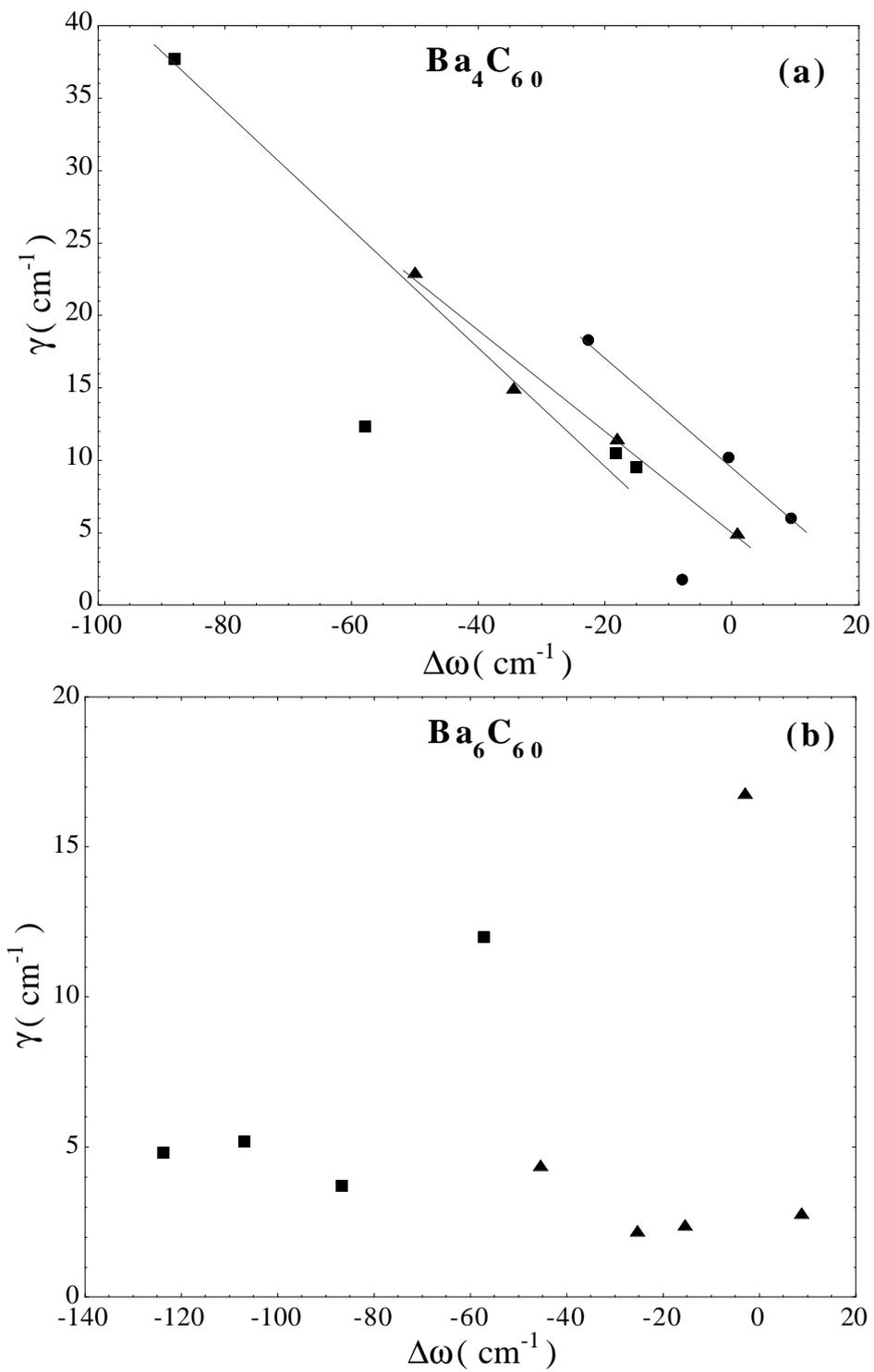
Figure 5:

Plot of linwidth $\gamma$ versus observed frequency shift $\Delta \omega $ for
the individual  components  of  the $Hg_(1)$, $H_g(2)$, and $H_g(3)$  modes,
circles  for  the $H_g(1)$ mode; triangles  for the $H_g(2)$ mode;  squares
for the $H_g(3)$ mode. (a) for the sample  $Ba_4C_{60}$;  (b) for the sample
$Ba_6C_{60}$.\\

\noindent
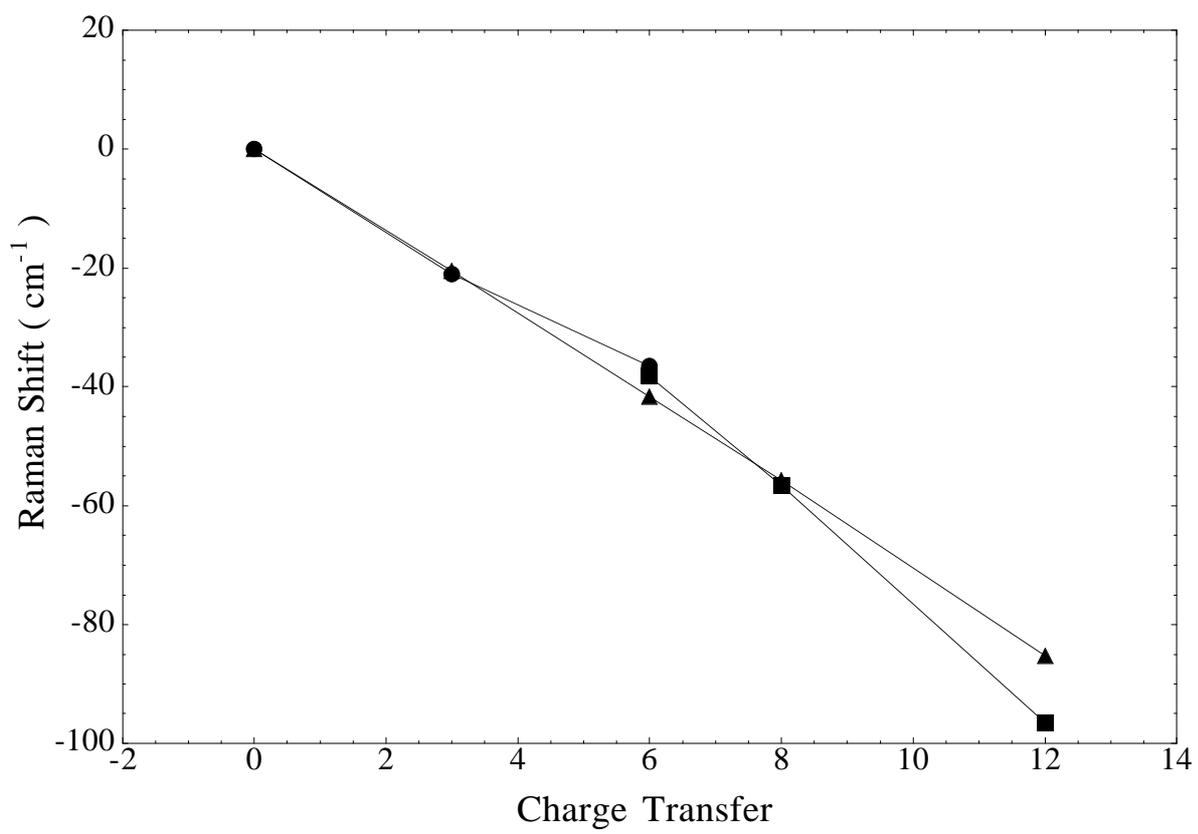
Figure 6:

Charge  transfer-Raman shift relation for the $A_g(2)$ pinch mode. Squares
 represent  the experimental  results of $Ba_xC_{60}$,  circles  are  from the 
results  of $K_xC_{60}$  reported  by  Duclos  et al. (Ref.15), and triangles  
refer to calculations  from theory of  Jishi and Dresselhaus  (Ref.16).

\end{document}